# Agile UAV landing control on moving ship in adverse conditions

James Mordaunt [1], Xinhua Wang [1†]
1 Aerospace Engineering, University of Nottingham, UK
† Email: wangxinhua04@gmail.com

**ABSTRACT**
This paper presents an agile Unmanned Aerial Vehicle (UAV) landing control by considering the effect of ship's oscillations and moving, and also disturbance (i.e., crosswind) is considered. The presented control system can make the quadrotor UAV autonomously land whilst overcoming these adverse conditions, and the addition of a rudder beneath each propeller is designed to increase the yaw authority which is found to be lacking in heavy-lift quadrotor UAV. The PID flight control system is proposed based on reference-point tracking, allowing the UAV to follow any desired path in 3D space whilst simultaneously yawing to face any desired heading. Realistic saturation limits on actuator outputs to ensure the real-world performance of actuators. Disturbances include randomised gusting wind in 3 axes, and sensor noise on translation and rotation signals to represent noise from the GPS and accelerometer respectively. The results from the simulations demonstrate that the UAV is capable of landing on a ship which is moving with varying heading and oscillating vertically on ocean waves and has the ability to time its descent such that it meets the ship at the peak of a wave to minimise the relative velocity.

**NOMENCLATURE**

| Symbol | Name | Unit |
|---|---|---|
| $m$ | Mass | $kg$ |
| $g$ | Gravitational acceleration | $ms^{-2}$ |
| $x_w, x_w, x_w$ | World-frame axes | – |
| $x_b, x_b, x_b$ | Body-frame axes | – |
| $\theta, \phi, \psi$ | Euler rotation angles | $rad$ |
| $J_\theta, J_\phi, J_\psi$ | Moments of inertia | $kg \cdot m^2$ |
| $T$ | Propeller thrust | $N$ |
| $\tau$ | Propeller torque | $Nm$ |
| $u_z, u_x, u_y$ | Propeller force in body axes | $N$ |
| $u_\theta, u_\phi, u_\psi$ | Propeller torque in body axes | $Nm$ |
| $K$ | Propeller force constant | $N/(rad \cdot s^{-1})$ |
| $K_\psi$ | Propeller torque constant | $m$ |
| $K_r$ | Rudder force constant | – |
| $K_p, K_i, K_d$ | Controller gains | – |
| $\omega_c$ | Filter cut-off frequency | $rad \cdot s^{-1}$ |
| $\omega$ | Angular velocity | $rad \cdot s^{-1}$ |
| $L$ | Arm length | $m$ |
| $A$ | Cross-sectional area | $m^2$ |
| $C_d$ | Drag coefficient | – |
| $\rho$ | Air density | $kg \cdot m^{-3}$ |

## 1 INTRODUCTION

Operating helicopters from ships presents a danger to both crew and vehicle, with risks greatly increasing in poor weather due to wind disturbance and ship oscillations caused by ocean waves. Helicopters take off and land on ships at sea for a wide range of critical civil and military missions [1], [2]. These missions may need to take place regardless of environmental conditions, meaning that a helicopter may be required to operate from a ship in adverse weather; this causes movement in the ship from ocean waves as well as wind disturbances on the aircraft itself [3], greatly increasing the risk of mishap to the aircraft, ship, and crew.
This paper proposes replacing manned helicopters with heavy-lift quadrotor UAVs where possible, reducing the aforementioned risks and decreasing pilot workload [1], [4].

The aim of this paper is to develop a control system to land a quadrotor UAV on a moving ship whilst experiencing wind disturbance and ship oscillations. The presented control system can make the quadrotor UAV autonomously land whilst overcoming these adverse conditions, and the addition of a rudder beneath each propeller is designed to increase the yaw authority which is found to be lacking in heavy-lift quadrotor UAV. This addition is evaluated using Computational Fluid Dynamics (CFD). The control system is based on reference-point tracking, allowing the UAV to follow any desired path in 3D space whilst simultaneously yawing to face any desired heading. The control system and UAV, ship, and disturbance models have all been simulated and evaluated in MATLAB Simulink. Realistic saturation limits on actuator outputs are used to ensure an accurate representation of real-world performance. Disturbances include randomised gusting wind in 3 axes, and sensor noise on translation and rotation signals to represent noise from the GPS and accelerometer respectively. The results from the simulations demonstrate that the UAV is capable of landing on a ship which is moving with varying heading and oscillating vertically on ocean waves and has the ability to time its descent such that it meets the ship at the peak of a wave to minimise the relative velocity. At touchdown, the position error was $0.110\,m$ and the relative velocity was $0.963\,ms^{-1}$, compared to a target relative velocity of $1\,ms^{-1}$, indicating a performance sufficient to replace manned helicopters for appropriate missions.

## 2 BACKGROUND

### 2.1 QUADROTOR CONTROL

The typical structure of a quadrotor is shown in Figure 1.

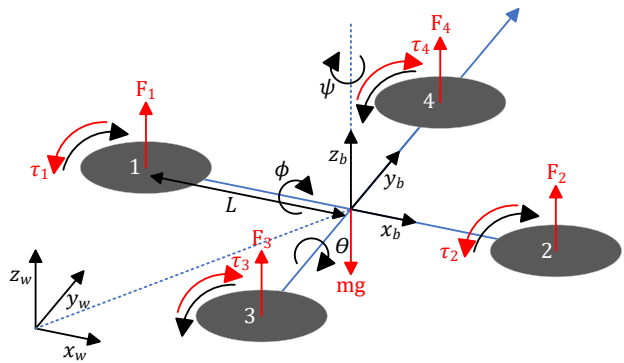

**Figure 1: Coordinate system and forces**

All 4 motors and propellers on a quadrotor act in the same fixed axis relative to the body, providing the benefit of increased payload capacity [5] compared to a helicopter, on which thrust is generated in 2 perpendicular axes. However, this results in the quadcopter being an under-actuated system as the number of degrees of freedom is greater than the number of axes in which the actuators can operate.

To overcome this limitation, the control of the quadcopter operates as follows: the force on the body is the sum of forces of the 4 propellers, i.e., $\sum T_i$; propellers on adjacent corners turn in opposite directions to give a net torque of 0, i.e., $\sum \tau_i = 0$; yaw can be controlled by increasing thrust on an opposite pair whilst decreasing thrust on the other opposite pair, such that $\sum T_i$ remains the same but $\sum \tau_i \neq 0$; pitch and roll are controlled by increasing the relative force on a pair of adjacent propellers whilst decreasing the relative force on the opposite pair, giving a torque across the body [6].

Translation of the body is achieved by holding a rotation angle in the desired direction and applying a thrust force. When nearing the desired position, the controller must command an attitude opposite to that which enabled its current position rate, providing a deceleration to prevent overshoot.

The controller must be capable of performing this in 3 axes simultaneously [7].

### 2.1.1 YAW AUTHORITY

The torque produced by each propeller is a combination of air resistance on the blades and recoil from the propeller's acceleration [8], which is of the form:

$$\tau = J\dot{\omega} \tag{1}$$

where $\tau$ is the recoil torque, $J$ is the moment of inertia, and $\dot{\omega}$ is angular acceleration. As the mass of the quadrotor increases, the size and therefore mass of its propellers must too, resulting in large quadrotor UAVs having poor yaw authority. This paper will introduce a new concept to help address this shortcoming (see Section 3.1.1).

## 2.2 EULER ANGLES

The rotation of the quadrotor's body relative to the world-frame is described by Euler angles $[\theta\ \phi\ \psi]$ in this paper, as is convention [7], [9]. An example can be seen in Figure 2. Further reading is recommended if this concept is not familiar to the reader, however, for simplicity, they can also be considered as roll, pitch, and yaw respectively.

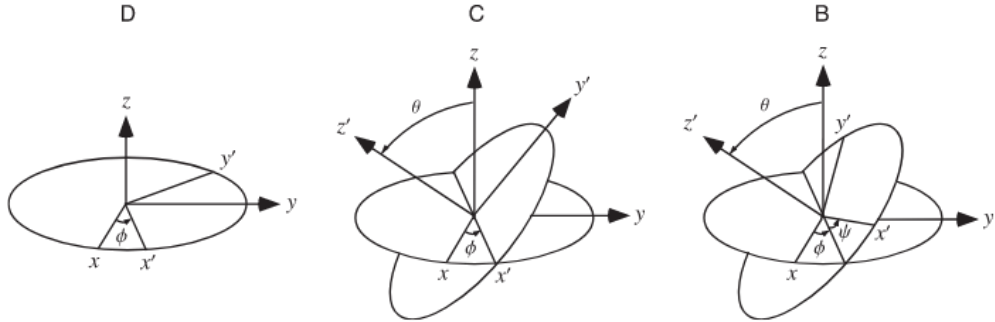

**Figure 2: A sequence of Euler rotations on a body** [10]

## 2.3 PID CONTROL

Proportional, Integral, and Derivative (PID) control offers a robust method of controlling complex dynamic systems [11]. It relies on calculating the error between a parameter's setpoint and current value from the plant to inform its output. A block diagram of the PID controller is shown in Figure 3. The performance of the controller can be tuned by adjusting the values of $K_p$, $K_i$, and $K_d$; $K_p$ increases the output proportionally to the error, $K_i$ integrates the error over time to reduce residual error, and $K_d$ checks the rate at which the error is changing to reduce overshoots.

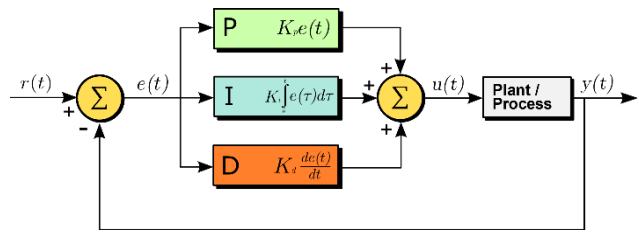

**Figure 3: PID controller block diagram** [12]

The PID controller can be written mathematically, as shown in Equation (2):

$$u_n(t) \stackrel{\text{def}}{=} K_{np}e_n(t) + K_{ni}\int_0^t e_n(\tau)d\tau + K_{nd}\frac{de_n(t)}{dt} \tag{2}$$

## 2.4 SENSOR NOISE AND FILTERING

All real-world sensors are flawed in that they add noise to the measured signal [9]. The ratio of noise to signal will vary depending on the amplitude of the signal and the sensitivity of the instrument, but for the purposes of this paper is assumed to not exceed $1:10$.
A control system's ability to filter and resist noise is an important measure of its capability. In this paper, noise is modelled by adding high-frequency random numbers to the outputs of the dynamic models before the signal is passed to the controller to be filtered. Different noise parameters are applied to the translational and rotational position controllers respectively to reflect the different types of sensors used for each.

The filtering can be performed by a first-order transfer function [13], which is of the form:

$$\frac{\omega_c}{s + \omega_c} \tag{3}$$

where $\omega_c$ is the cut-off frequency. The value of $\omega_c$ is optimised in Section 4.3. Figure 4 shows how a well-optimised filter is able to produce an output close to the original signal before noise injection.

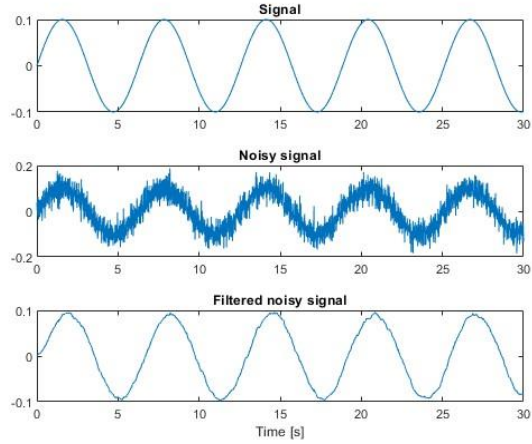

**Figure 4: Demonstration of the filtering performance of a first-order transfer function with $\omega_c = 2.5\ rad/s$**

## 3    SYSTEM DESCRIPTION

### 3.1  QUADROTOR STRUCTURE

The coordinate system and motor numbers for the quadrotor modelled in this paper has been developed and shown in Figure 5.

The physical parameters of the quadrotor being considered in this paper are listed in Table 1. For the purpose of simulation in Simulink, these values are defined in a MATLAB script so can quickly be modified by the user if desired.

**Table 1: Quadrotor physical parameters**

| Variable | Value | Unit |
|:---:|:---:|:---:|
| $m$ | 100 | $kg$ |
| $J_\theta$ | 0.1 | $kg \cdot m^2$ |
| $J_\phi$ | 0.1 | $kg \cdot m^2$ |
| $J_\psi$ | 0.01 | $kg \cdot m^2$ |
| $K$ | 1 | $N/(rad \cdot s^{-1})$ |
| $K_\psi$ | 0.5 | $m$ |
| $\omega_{max}$ | 500 | $rad \cdot s^{-1}$ |
| $L$ | 0.75 | $m$ |
| $A$ | 1 | $m^2$ |
| $C_d$ | 0.5 | $-$ |

#### 3.1.1 RUDDER STRUCTURE

As discussed in Section 2.1.1, heavy-lift quadrotors have poor yaw authority. This paper proposes the addition of a rudder beneath each propeller to increase yaw authority and thus the dynamic control of the vehicle. The traditional quadrotor structure, shown in Figure 1, has been modified, as shown in Figure 5. The sophistication of this arrangement should be significantly refined for any real-world implementation; however, a primitive model is sufficient at this stage. Note that the actuation mechanism is not shown.

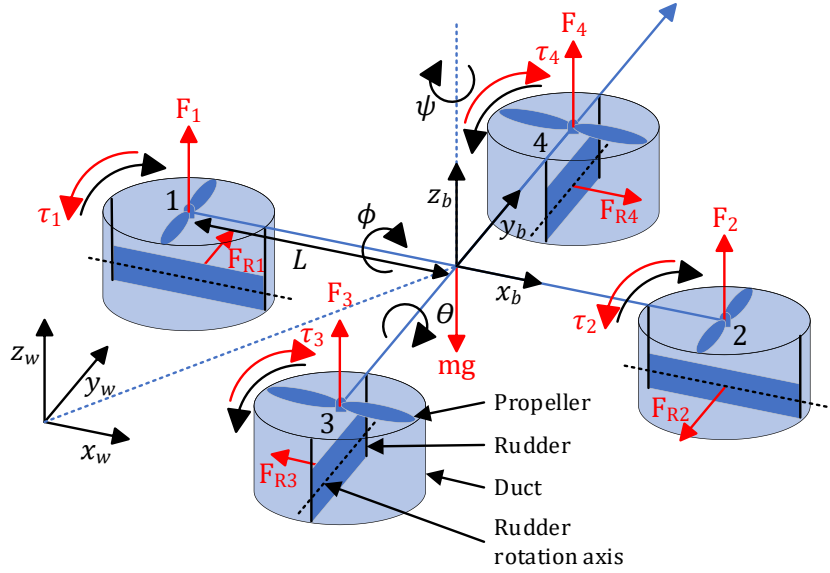

**Figure 5: Schematic view of propeller-rudder arrangement**

## 3.2  FORCES AND TORQUES

The relationships between propeller forces and body forces have been derived from Qiao [6] in equations (4)-(7), along with the coordinate system and motor numbers shown in Figure 5:

$$u_z = T_1 + T_2 + T_3 + T_4 \tag{4}$$

$$u_\theta = L(T_1 - T_3) \tag{5}$$

$$u_\phi = L(T_2 - T_4) \tag{6}$$

$$u_\psi = \tau_1 + \tau_2 + \tau_3 + \tau_4 \tag{7}$$

where $u_n$ is the force or torque in direction $n$, $T_i$ is the thrust produced by motor $i$, $L$ is the arm length (i.e., the distance from the centre of mass to the motor), and $\tau_i$ is the torque produced by motor $i$, where,

$$\tau_i = K_\psi T_i \tag{8}$$

$$T_i = K u_i \tag{9}$$

where $K_\psi$ and $K$ are constants, and $u_i$ is the speed of motor $i$.

Equations (4)-(9) can then be re-written in matrix form as:

$$\begin{bmatrix} u_z \\ u_\theta \\ u_\phi \\ u_\psi \end{bmatrix} = \begin{bmatrix} K & K & K & K \\ KL & -KL & 0 & 0 \\ 0 & 0 & KL & -KL \\ KK_\psi & KK_\psi & -KK_\psi & -KK_\psi \end{bmatrix} \begin{bmatrix} u_1 \\ u_2 \\ u_3 \\ u_4 \end{bmatrix} \tag{10}$$

Equation (10) will later be rearranged to calculate motor speeds based on the desired force outputs from the controller.

## 3.3 POSITION DYNAMICS

A model for the position dynamics of the quadrotor has been adapted from Qiao [6] in equations (11)-(13):

$$m\ddot{x} = u_z(\cos\phi \sin\theta \cos\psi + \sin\phi \sin\psi) + F_{x\,wind} + \delta_{x\,noise} \tag{11}$$

$$m\ddot{y} = u_z(\cos\phi \sin\theta \sin\psi - \sin\phi \cos\psi) + F_{y\,wind} + \delta_{y\,noise} \tag{12}$$

$$m\ddot{z} = u_z(\cos\phi \cos\theta) - mg + F_{z\,wind} + \delta_{z\,noise} \tag{13}$$

where $m$ is the mass of the quadrotor, $\ddot{n}$ is the acceleration of the quadrotor in the world-frame axis $n$, $u_z$ is the force in the body-frame $z$ direction from the motors (see equation (4)), $g$ is gravitational acceleration, $F_{n\,wind}$ is the aerodynamic force on the body in world-frame axis $n$ (see Section 3.3.1), $\delta_{n\,noise}$ is a disturbance in the measured value of $n$ as a result of sensor noise (see Section 2.4), and $[\theta\,\phi\,\psi]$ are the quadrotor's Euler rotation angles (see Section 2.2).

### 3.3.1 WIND MODEL

A model for the behaviour of wind and its force on the quadrotor has been derived from Tran [14], Jeon [15], and Viktor [16]:

$$F_{n\,wind}(t) = K_d\big(\dot{n}(t) - wind_n(t)\big)^2 \big(-sign(\dot{n}(t) - wind_n(t))\big) \tag{14}$$

where,

$$K_d = \frac{1}{2} C_d \rho A \tag{15}$$

$$wind_n(t) = steady_n + gust_n(t) \tag{16}$$

$$steady_n = rand(wind_{min}, wind_{max}) * rand[-1\,1] \tag{17}$$

$$gust_n(t) = \frac{steady_n}{5} white\_noise\big(t, (-1,1)\big) \tag{18}$$

Equation (14) is a slight modification of the standard drag equation; $\dot{n} - wind_n$ is the relative velocity of the quadcopter to the wind in axis $n$ (also referred to as indicated airspeed), and the $-sign$ function modifies to formula to make the force act in the opposite direction to positive relative velocity, i.e., as a drag force.
Equation (16) simply describes the wind force as the sum of the steady wind and the wind gust. Equation (17) shows that the steady wind is randomly selected from within a user-specified range and is then randomly assigned to be positive or negative in that axis. Equation (18) describes the generation of wind gusts as random velocities in the range of $-20\%$ to $+20\%$ of the steady wind velocity. The white noise sample time $t$ determines the rate at which the gust changes and is user-specified.

## 3.4 ATTITUDE DYNAMICS

The attitude dynamics of the quadrotor have been derived and are as follows:

$$J_\theta \ddot{\theta} = u_\theta + \delta_{\theta\,noise} \tag{19}$$

$$J_\phi \ddot{\phi} = u_\phi + \delta_{\phi\,noise} \tag{20}$$

$$J_\psi \ddot{\psi} = u_\psi K_r + \delta_{\psi\,noise} \tag{21}$$

where $J_n$ is the moment of inertia about axis $n$, $\ddot{n}$ is angular acceleration about axis $n$, $u_n$ is the torque about axis $n$ (see Section 3.2), $K_r$ is a constant describing the rudder force derived from the simulations discussed in Section 4.1, and $\delta_{n\,noise}$ is a disturbance in the measured value of $n$ as a result of sensor noise (see Section 2.4).

## 3.5 SHIP MODEL

A simple model has been designed and implemented. The ship is modelled as a point in 3D space. It oscillates vertically sinusoidally with user-specified phase, frequency, and amplitude. It translates simultaneously in the horizontal plane with user-specified starting position, heading, and velocity, with the latter two parameters able to change dynamically during the simulation.

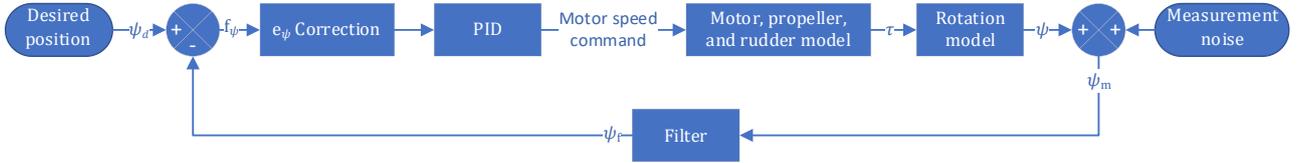

**Figure 6: Control loop layout for $\psi$**

Whilst far more sophisticated models are available and documented [2], [17]–[19], the focus of this paper is on the development of a reference-point tracking control system so a simple approximation is sufficient.

A more sophisticated model may account for the variation in wave types across the world, variable velocity and amplitude waves, and the combination of waves from different sources to produce a complex multi-dimensional problem.

## 3.6 CONTROLLER DESIGN

### 3.6.1 FLIGHT CONTROLLER

A flight controller has been designed which allows the UAV to fly to and follow a stationary or moving desired position in 3D space, and simultaneously yaw to face any desired heading. The design of the flight controller is inspired by Selby [20], Bouabdallah [7], and Sawyer [9].

To determine the required speed of each motor, Equation (10) can be rearranged as:

$$\begin{bmatrix} u_1 \\ u_2 \\ u_3 \\ u_4 \end{bmatrix} = \begin{bmatrix} K & K & K & K \\ KL & -KL & 0 & 0 \\ 0 & 0 & KL & -KL \\ KK_\psi & KK_\psi & -KK_\psi & -KK_\psi \end{bmatrix}^{-1} \begin{bmatrix} u_z \\ u_\theta \\ u_\phi \\ u_\psi \end{bmatrix} \quad (22)$$

where the $4 \cdot 4$ matrix is constant and pre-computed. The controllers for $u_z$, $u_\theta$, $u_\phi$, and $u_\psi$ will now be defined.

For each of the following controller equations,

$$e_n(t) = n_d(t) - n(t) \quad (23)$$

where $n(t)$ is the value of parameter $n$ at time $t$ and $n_d(t)$ is the desired value of that parameter at time $t$.

As discussed in Section 2.1, the direction of the quadrotor's thrust is in the $z$ direction of the body-frame. The controller for this thrust in the body-frame is shown in equation (24):

$$u_{z_b}(t) \stackrel{\text{def}}{=} K_{zp} e_z(t) + K_{zi} \int_0^t e_z(\tau) d\tau + K_{zd} \frac{de_z(t)}{dt} \quad (24)$$

A transformation can then be applied to calculate the thrust force in the $z$ direction of the world-frame:

$$u_z(t) = \frac{u_{z_b}(t)}{\cos\theta(t)\cos\phi(t)} \quad (25)$$

The controllers shown in equations (26) and (27) determine the desired force in the $x$ and $y$ world-frame directions:

$$u_x(t) \stackrel{\text{def}}{=} K_{xp}e_x(t) + K_{xi}\int_0^t e_x(\tau)d\tau + K_{xd}\frac{de_x(t)}{dt} \quad (26)$$

$$u_y(t) \stackrel{\text{def}}{=} K_{yp}e_y(t) + K_{yi}\int_0^t e_y(\tau)d\tau + K_{yd}\frac{de_y(t)}{dt} \quad (27)$$

A transformation can then be applied to account for the difference between world and body-frames when the quadrotor is facing some direction other than North, i.e., $\psi \neq 0$, allowing the desired pitch and roll values to be determined:

$$\theta_d(t) = (u_x(t) * \cos\psi(t)) + (u_y(t) * \sin\psi(t)) \quad (28)$$

$$\phi_d(t) = (u_x(t) * \sin\psi(t)) + (u_y(t) * -\cos\psi(t)) \quad (29)$$

Finally, the controllers shown in equations (30) and (31) determine the torques about the pitch and roll axes:

$$u_\theta(t) \stackrel{\text{def}}{=} K_{\theta p}e_\theta(t) + K_{\theta i}\int_0^t e_\theta(\tau)d\tau + K_{\theta d}\frac{de_\theta(t)}{dt} \quad (30)$$

$$u_\phi(t) \stackrel{\text{def}}{=} K_{\phi p}e_\phi(t) + K_{\phi i}\int_0^t e_\phi(\tau)d\tau + K_{\phi d}\frac{de_\phi(t)}{dt} \quad (31)$$

Care should be taken in the design of the yaw controller, shown in

$$u_\psi(t) \stackrel{\text{def}}{=} K_{\psi p}e_\psi(t) + K_{\psi i}\int_0^t e_\psi(\tau)d\tau + K_{\psi d}\frac{de_\psi(t)}{dt} \quad (32)$$

$$e_\psi(t) = \begin{cases} (f_\psi \bmod 2\pi) + 2\pi, & f_\psi \bmod 2\pi < -\pi \\ f_\psi \bmod 2\pi, & -\pi \leq f_\psi \bmod 2\pi \leq \pi \\ (f_\psi \bmod 2\pi) - 2\pi, & \pi < f_\psi \bmod 2\pi \end{cases} \quad (33)$$

$$f_\psi = \psi_d - \psi \quad (34)$$

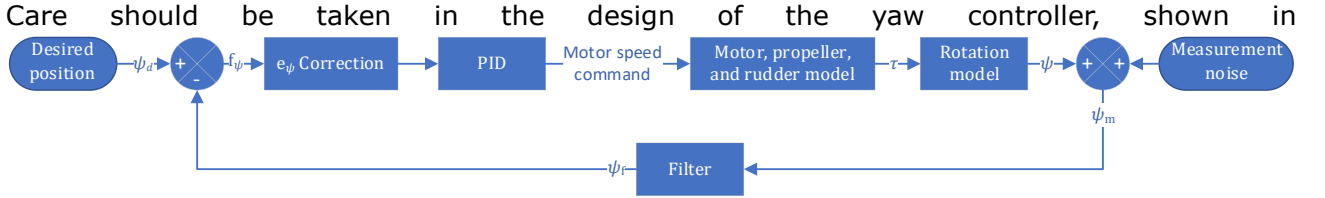

Figure 6. Whilst the PID controller takes the same form as normal, the definition of error $e_\psi(t)$ changes depending on the current state:

To explain this logic, consider the situation shown in Figure 7; if the UAV is currently following Heading 1 and the mission profile requires it to change to Heading 2, an uncorrected controller would command an increase in heading to move from 005° to 345°, where increasing heading equates to clockwise rotation when viewed from above, i.e., the red path. It is clear that the UAV should take the shorter green path instead.

To correct for this, the controller first performs a $mod$ function on the error to negate the effect of multiple previous rotations, and output only the current error within the period of one rotation. This result is then checked against the angle of a half-rotation to determine whether it would be more efficient to assume it to be within the next or previous rotational period, rather than the current, and if so, the appropriate adjustment is applied. This process is shown in equations (33) and (34), and explained diagrammatically in Figure 8.

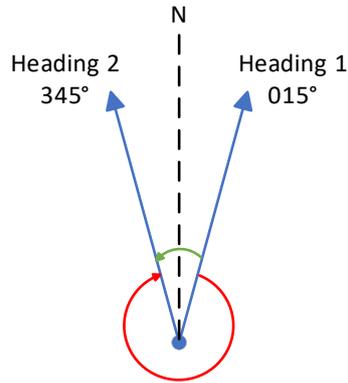

**Figure 7: The two options for crossing North shown by green and red arrows**

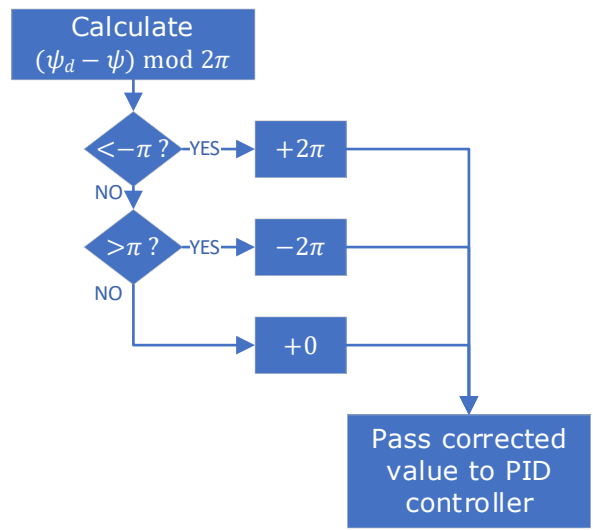

**Figure 8: Flowchart showing the method to correct $\psi$ error to ensure efficient North crossing**

### 3.6.2 SATURATION LIMITS

Saturation limits have been put in place in the controller.

In the case of $\theta_d$ and $\phi_d$, the saturation limit is the maximum rotation angle that the controller should command in order to maintain stable flight. Without this limit, a position error in $x$ or $y$ would cause the controller to command continuously increasing values of $\theta$ and $\phi$, placing the quadrotor into an uncontrolled spin.

The saturation limits on the outputs of the controller, i.e., $u_z$, $u_\theta$, $u_\phi$, and $u_\psi$ are in place to limit the performance of the quadrotor to realistic capabilities. The limit values, shown in Table 2, are derived from the force and torque calculations discussed in Section 3.2, as suggested by Selby [20], combined with a new term $\omega_{max}$ which is a user-specified maximum angular velocity for the motor-propeller system.

**Table 2: Controller parameter saturation limits**

| Parameter | Minimum | Maximum |
|---|---|---|
| $\theta_d, \phi_d$ | $-\frac{\pi}{4}$ | $\frac{\pi}{4}$ |
| $u_z$ | 0 | $4K\omega_{max}$ |
| $u_\theta, u_\phi$ | $-LK\omega_{max}$ | $LK\omega_{max}$ |
| $u_\psi$ | $-2KK_\psi \omega_{max}$ | $2KK_\psi \omega_{max}$ |

### 3.6.3 LANDING CONTROLLER

A landing controller has been developed to control the descent of the UAV from its holding altitude to the ship.

The function of the landing controller is to time the descent of the quadrotor such that it comes into contact with the ship when the ship is at the peak of a wave, minimising the relative velocity to reduce the risk of damage. The controller has been summarised in Figure 9.

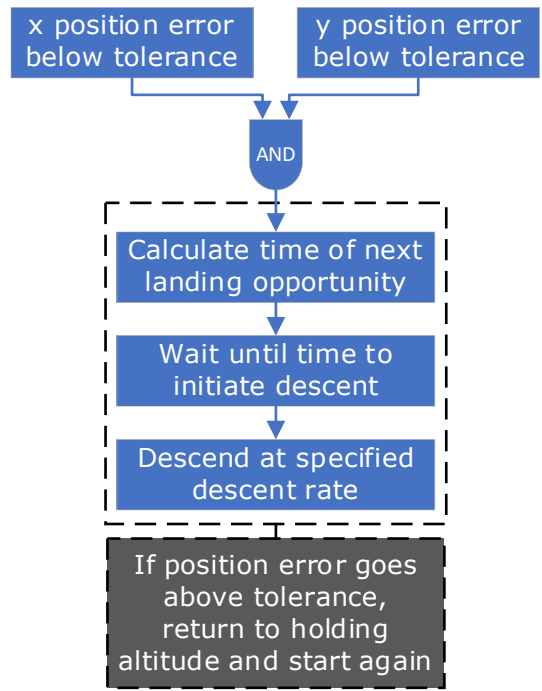

**Figure 9 : Flowchart showing the logic within the landing controller**

## 4  RESULTS

### 4.1  RUDDER CONCEPT CFD

The rudder concept has been analysed using CFD and demonstrated to be feasible.
The specific aims of this CFD analysis are:

1. to assess whether the force on the rudder should be modelled as periodic or constant,
2. to determine the loss of thrust as a result of drag on the rudder, and
3. to collect data from which an empirical model can be generated.

A suitably refined propeller model was found [21] and combined in SolidWorks with a rudder and enclosure. A 3D, transient, moving mesh analysis was performed on the propeller-rudder arrangement using Ansys Fluent, a software package recognised and accepted by industry. The analysis was run with rudder angles of $0°$, $5°$, $10°$, and $15°$, and propeller speeds of $200$, $300$, and $400 \, rad \cdot s^{-1}$ at each rudder angle. Following a sensitivity analysis, the mesh consisted of 1,631,700 elements and each case took approximately 9 hours to complete 0.5 seconds of simulation. Figure 10 shows an example of the simulation outputs during testing.

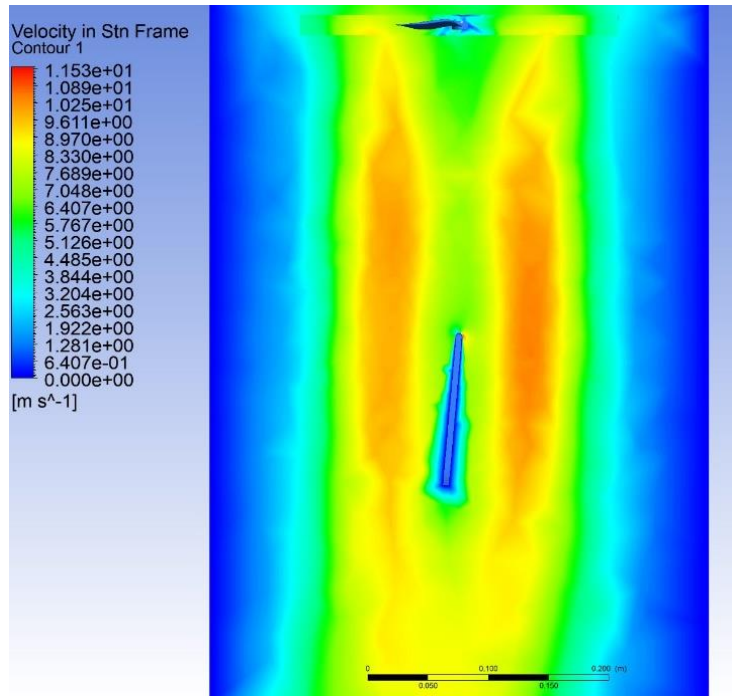

**Figure 10: The flow of air generated by the propeller over the rudder during simulation**

The simulation shows that the rudder does not experience a periodic force, so can be modelled as constant. The arrangement maintains a net thrust of $96.2\%$ of gross thrust in the worst case (rudder angle of $15°$, propeller speed of $200\ rad \cdot s^{-1}$), demonstrating an acceptable loss from drag on the rudder. Figure 11 shows the change in horizontal force on the rudder in each case.

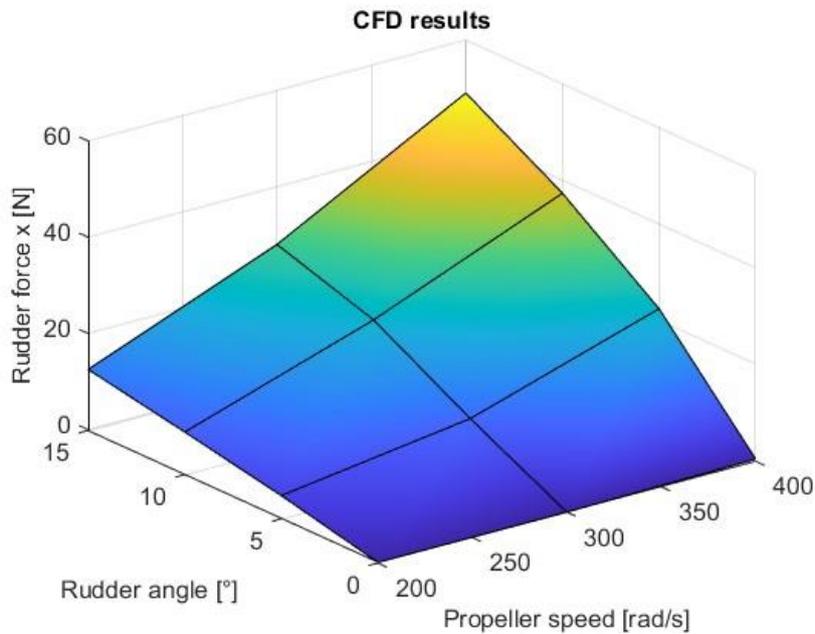

**Figure 11: Horizontal force on rudder as a function of rudder angle and propeller speed**

This analysis concludes that the addition of a rudder beneath each propeller on a quadrotor UAV is a feasible method of increasing yaw authority.

### 4.2 PID GAIN TUNING

The gains for the controllers have been tuned to give a desirable response.

There are 4 PID controllers and 2 PD controllers in the model. As discussed in Section 2.3, each of these requires gain tuning to ensure an appropriate response and robust control.

Simulink provides a tool for transfer-function based automated tuning of PID controllers [22]. This was used successfully to determine values for the lowest-hierarchy controllers, i.e., those whose outputs do not feed into another controller: $u_z$, $u_\theta$, $u_\phi$, and $u_\psi$. For the controllers of $u_x$ and $u_y$, manual tuning was performed based on an understanding of the effects of each parameter, discussed in Section 2.3. The PID gains used in the final model are shown in Table 3.

**Table 3: PID gains**

| Controller output | P | I | D |
|---|---|---|---|
| $u_z$ | 297.08 | 55.6 | 389 |
| $u_\theta$ | 0.324 | − | 0.383 |
| $u_\phi$ | 0.324 | − | 0.383 |
| $u_\psi$ | 0.00485 | $3.83e-5$ | 0.0518 |
| $u_x$ | 0.500 | 0.0500 | 0.500 |
| $u_y$ | 0.500 | 0.0500 | 0.500 |

## 4.3 FILTER COEFFICIENT OPTIMISATION

The optimal values of $\omega_c$ have been found for both the rotation and the translation signal filters.

In order to find the optimal value of $\omega_c$ (see Section 2.4), a 2D parameter sweep of $\omega_c = (0, 90)$ was conducted. The average 3D position error over the final 10% of the simulation time for a ship-landing mission was used as a performance metric, as shown in Figure 12. The plot has been limited to $3\,m$ on the $y$ axis, as values greater than this indicate a failure.

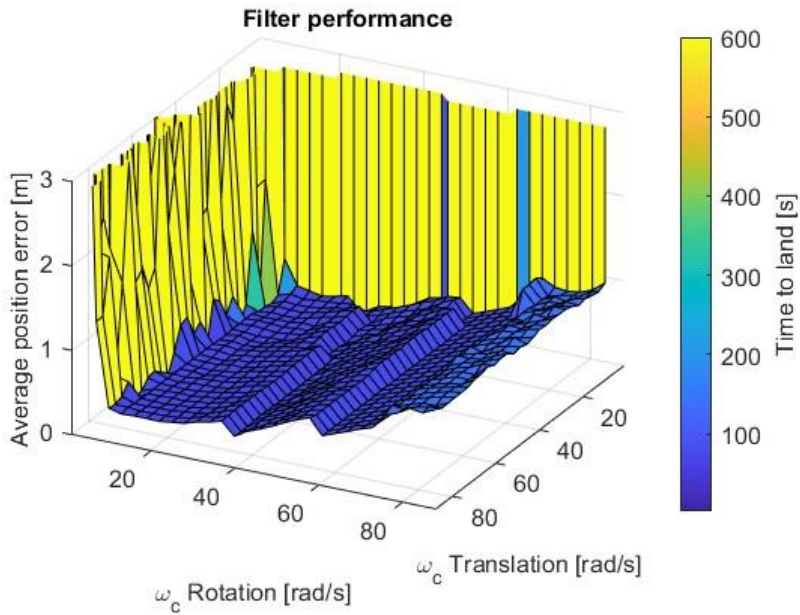

**Figure 12: The results of a parameter sweep searching for the optimal $\omega_c$ values**

Time to land has inherent "steps" in its values as the controller will always wait to land at the peak of a wave. A time to land of $600\,s$ indicates that the UAV failed to land, as this is the stop time of the simulation.

The testing shows that the optimal values of $\omega_c$ are $\omega_c = 13\,rad/s$ for the translation signal filter and $\omega_c = 31\,rad/s$ for the rotation signal filter.

## 4.4 PERFORMANCE

The overall performance of the system has been assessed in 3 different scenarios: following a Lissajous curve, following a spiral climb, and hovering in place. Each scenario was run both with and without wind, however only one figure for each scenario has been presented in this paper. Sensor noise was present in all scenarios.

The Lissajous path shown in Figure 13 demonstrates the ability to change direction and follow curves of variable radius. The consistency of the UAV's path can be seen in the repeated circuits of the curve, and performance is similar to the examples produced by Sawyer [9].

An initial instability can be seen. This is caused by the UAV beginning at $[x\ y\ z] = [\dot{x}\ \dot{y}\ \dot{z}] = 0$, so it must accelerate to catch the moving reference point which does not have an acceleration period. Additionally, the UAV initially loses altitude as the controller ramps up its output to reach a stable hover. This cannot be seen in this 2D plot.

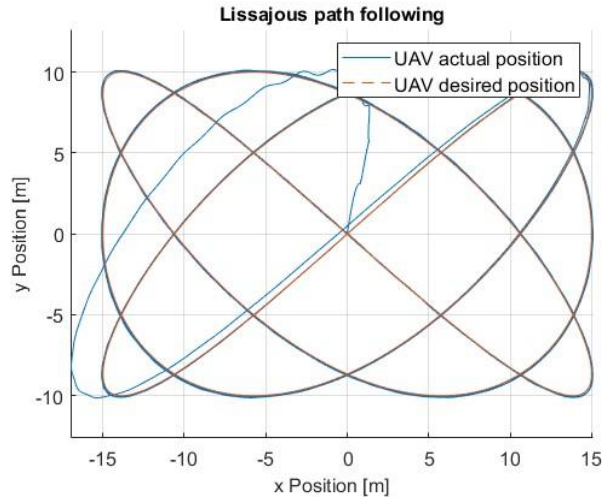

**Figure 13: Controller performance in following a Lissajous curve in a windless environment**

The spiral climb shown in Figure 14 demonstrates the ability to follow a repetitive track whilst also changing altitude, and again is similar to examples produced by Sawyer [9]. The same initial instability can be seen as with the previous example.

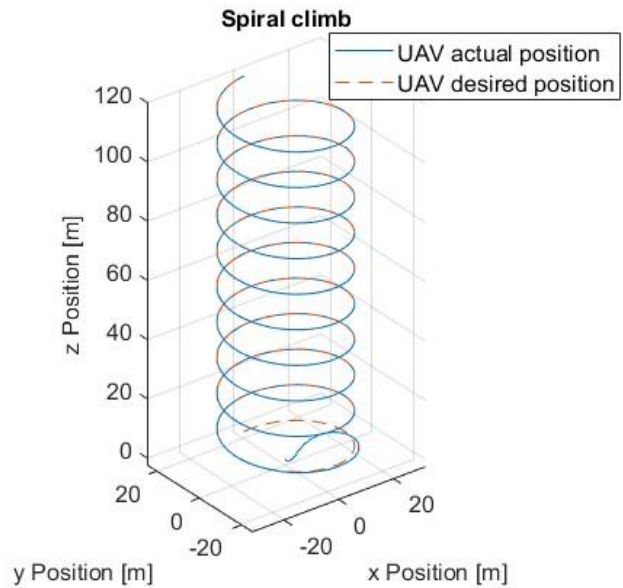

**Figure 14: Controller performance in following a spiral climb in a windless environment**

Finally, the hover shown in Figure 15 demonstrates the ability to hold position in the presence of random gusting wind in 3 dimensions. It shows marginally worse performance than in examples produced by Hu [23], however this study did not include sensor noise as a source of disturbance.

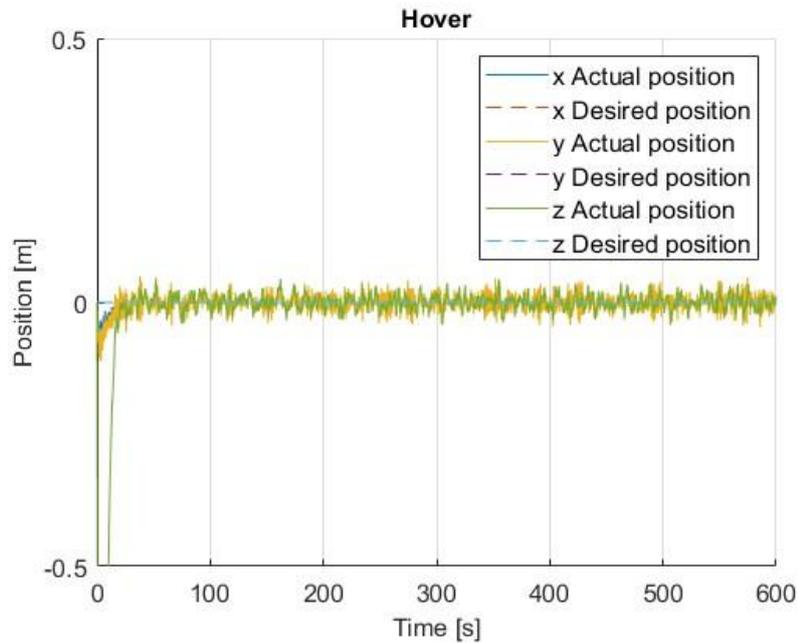

**Figure 15: Controller performance in holding position in a randomised wind environment**

The mean 3D position error in each scenario is shown in Table 4. Only data from after 25% of total simulation time has been included to exclude the initial instability at the start of each scenario.

**Table 4: Mean 3D position error for the final 75% of simulation time in each scenario [m]**

| Scenario | Wind ($10 - 20 \, ms^{-1}$) | No wind |
|---|---|---|
| Lissajous | 0.077 | 0.022 |
| Spiral | 0.055 | 0.053 |
| Hover | 0.048 | 0.013 |

### 4.4.1 LANDING CONTROL

A ship landing mission was designed and simulated using the parameters shown in Table 5.

Figure 16 shows the UAV performing the mission for which it was designed: landing on a moving ship. The UAV starts at a point in space and translates to align itself with the moving ship. Once its position error is within tolerance, it triggers the descent controller.

The performance of the descent controller can be seen in Figure 17. The UAV waits at its holding altitude of $20 \, m$ whilst it aligns itself in $x$ and $y$. The controller plans then executes a descent so that it meets the ship at the peak of a wave to minimise the relative velocity.

The final position error was $0.110 \, m$ and the relative velocity at touchdown was $0.963 \, ms^{-1}$, compared to a target relative velocity of $1 \, ms^{-1}$.

**Table 5: Simulation parameters**

| Variable | Value | Unit |
|---|---|---|
| Ship initial position $(x, y)$ | $(500, 300)$ | $m$ |
| Ship initial heading | 300 | ° |
| Ship turn rate | 2 | °/$s$ |
| Ship speed | 15 | $ms^{-1}$ |
| Wind speed $(min, max)$ | $(10, 20)$ | $ms^{-1}$ |
| Wave amplitude | 5 | $m$ |
| Wave frequency | 0.75 | $rad \cdot s^{-1}$ |
| Wave phase | 2.2 | $rad$ |
| Holding altitude | 20 | $m$ |
| Position error tolerance | 1 | $m$ |
| Translation sensor noise variance | 0.001 | − |
| Translation sensor noise sample time | 0.01 | $s$ |
| Rotation sensor noise variance | 0.0001 | − |
| Rotation sensor noise sample time | 0.01 | $s$ |

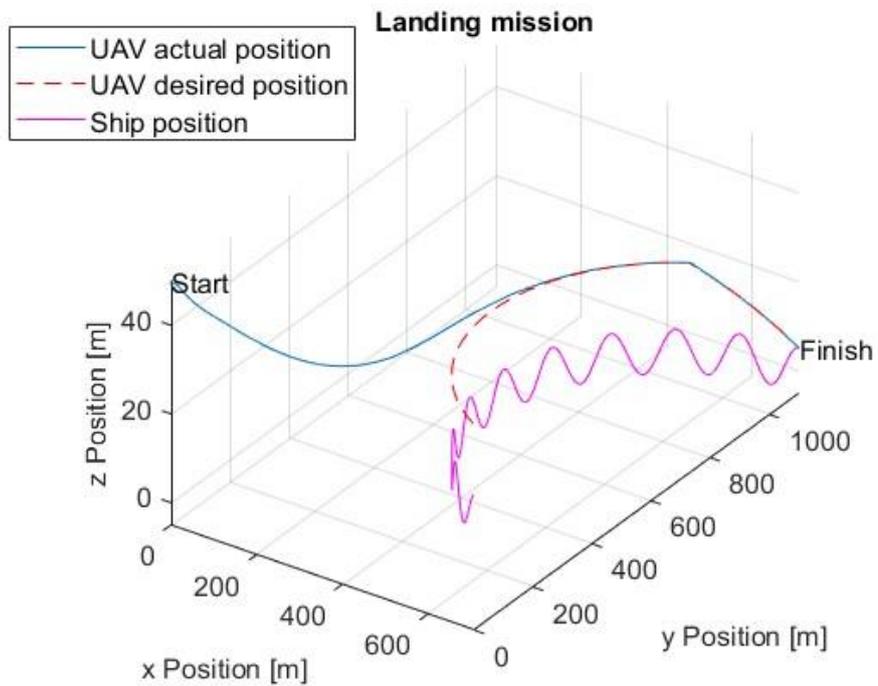

**Figure 16: UAV performing a landing on a moving ship**

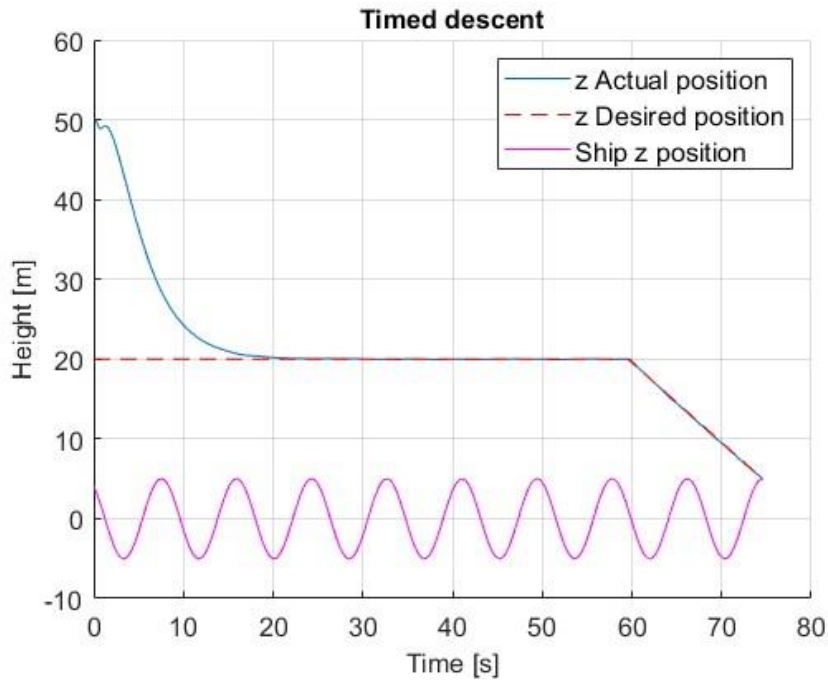

**Figure 17: Landing controller timing descent to meet the ship at the peak of a wave**

## 5   CONCLUSIONS

A control system has been developed to allow a quadrotor UAV to follow any path, including one which matches a moving and oscillating ship. The concept of the addition of a rudder beneath each propeller has been analysed and implemented. A simple model of the motion of a ship in waves has been produced, and the quadrotor's ability to follow a path to land on the ship has been assessed. A landing controller ensures that the position error is within tolerance and controls the timing and speed of the descent so that the quadrotor meets the oscillating ship at the peak of a wave, at a specified relative velocity. In testing, which included sensor noise and wind disturbance, the UAV landed on the ship with a position error of $0.110\ m$ and the relative velocity at touchdown was $0.963\ ms^{-1}$, compared to a target relative velocity of $1\ ms^{-1}$, indicating that this UAV as simulated has sufficient performance to replace manned helicopters for appropriate missions.